\font\tenfrakturb=eufb10
\font\tenfraktur=eufm10
\font\tenmsbm=msbm10
\font\sevenfrakturb=eufb7
\font\sevenfraktur=eufm7
\font\sevenmsbm=msbm7
\font\fivefrakturb=eufb5
\font\fivefraktur=eufm5
\font\fivemsbm=msbm5
\newfam\bgothicfam
\newfam\gothicfam
\newfam\msbmfam
\textfont\bgothicfam = \tenfrakturb \scriptfont\bgothicfam=\sevenfrakturb
\scriptscriptfont\bgothicfam=\fivefrakturb
\textfont\gothicfam = \tenfraktur \scriptfont\gothicfam=\sevenfraktur
\scriptscriptfont\gothicfam=\fivefraktur
\textfont\msbmfam = \tenmsbm \scriptfont\msbmfam=\sevenmsbm
\scriptscriptfont\msbmfam=\fivemsbm

\def\Bbb{\tenmsbm\fam\msbmfam}

\catcode`@=11
\def\renewcounter#1{\@definecounter{#1}\@ifnextchar[{\@newctr{#1}}{}}

\documentstyle{elsart}
\begin{document}
\begin{frontmatter}
\title{ Estimates for parameters and characteristics of the confining 
SU(3)-gluonic field in pions and kaons  }
\author{Yu. P. Goncharov}
\address{Theoretical Group, Experimental Physics Department, State Polytechnical
         University, Sankt-Petersburg 195251, Russia}

\begin{abstract} 
The estimates of parameters for the classical SU(3)-gluonic field responsible 
for linear confinement are given for pions and kaons. Also estimates for the 
characteristics of the mentioned field such as gluon concentrations, electric 
and magnetic colour field strengths are adduced along with those for quark 
velocities in mesons under discussion. Possible connection of the obtained 
results with potential approach and string-like picture of confinement is 
outlined as well. 
\end{abstract}
\end{frontmatter}   
\section{Introduction and preliminary remarks}

In Ref. \cite{Gon01} for the Dirac-Yang-Mills system derived from 
QCD-Lagrangian there was found a family of compatible nonperturbative solutions 
which could pretend to decsribing confinement of two quarks. 
Further study of this family as well as its applications to the quarkonia 
spectra (charmonium and bottomonium) in 
Refs. \cite{{Gon03},{GB04}} showed that the above family is in essence 
unique one \cite{Gon051} and allows one to modify gluon propagator in a 
nonperturbative way so that the modified propagator could correspond to linear 
confinement at large distances \cite{Gon052}. 

Two main physical reasons for linear confinement in the mechanism under 
discussion are the following ones. The first one is that gluon exchange between 
quarks is realized with the propagator different from the photon one and 
existence of such a propagator is {\em direct} consequence of the unique confining 
nonperturbative solutions of the Yang-Mills equations \cite{Gon051}. The second 
reason is that, 
owing to the structure of mentioned propagator, gluon condensate (a classical 
gluon field) between quarks mainly consists of soft gluons (for more details 
see \cite{{Gon051},{Gon052}}) but, because of that any gluon also emits 
gluons (still softer), the corresponding gluon concentrations 
rapidly become huge and form the linear confining magnetic colour field of 
enormous strengths which leads to confinement of quarks. Under the circumstances 
physically nonlinearity of the Yang-Mills equations effectively vanishes so the 
latter possess the unique nonperturbative confining solutions of the 
abelian-like form (with the values in Cartan subalgebra of SU(3)-Lie algebra) 
\cite{Gon051} that describe 
the gluon condensate under consideration. Moreover, since the overwhelming majority 
of gluons are soft they cannot leave hadron (meson) until some gluon obtains 
additional energy (due to an external reason) to rush out. So we deal with 
confinement of gluons as well. 

The approach under discussion equips us with the explicit wave 
functions that is practically unreachable in other approaches, for example, 
within framework of lattice theories or potenial approach. Namely, 
for each two quarks (meson) there exists its own set of real constants 
(for more details see below) $a_j, A_j, b_j, B_j$ parametrizing the mentioned 
{\em nonperturbative} confining gluon field (the gluon condensate) and the 
corresponding wave functions ({\em nonperturbative} modulo square integrable 
solutions of the 
Dirac equation in this confining SU(3)-field) 
while the latter ones also depend on $\mu_0$, the reduced
mass of the current masses of quarks forming 
meson. It is clear that constants $a_j, A_j, b_j, B_j,\mu_0$
should be extracted from experimental data. This circumstance gives 
possibilities for direct physical modelling of internal structure for any 
meson and for checking such relativistic models numerically. 

So far all applications of the confinement mechanism under consideration have 
been restricted to the quarkonia \cite{{Gon03},{GB04},{Gon051},{Gon052}}. The 
aim of the present paper is to estimate the above parameters 
for the case of pions and kaons. Of course, when conducting our considerations 
we should rely on the standard quark model (SQM) based on SU(3)-flavor symmetry 
(see, e. g., Ref. \cite{pdg} or the oldies \cite{RF}). 

Further we shall deal with the metric of
the flat Minkowski spacetime $M$ that
we write down (using the ordinary set of local spherical coordinates
$r,\vartheta,\varphi$ for the spatial part) in the form
$$ds^2=g_{\mu\nu}dx^\mu\otimes dx^\nu\equiv
dt^2-dr^2-r^2(d\vartheta^2+\sin^2\vartheta d\varphi^2)\>. \eqno(1)$$
Besides, we have $|\delta|=|\det(g_{\mu\nu})|=(r^2\sin\vartheta)^2$
and $0\leq r<\infty$, $0\leq\vartheta<\pi$,
$0\leq\varphi<2\pi$.

Throughout the paper we employ the Heaviside-Lorentz system of units 
with $\hbar=c=1$, unless explicitly stated otherwise, so the gauge coupling 
constant $g$ and the strong coupling constant ${\alpha_s}$ are connected by 
relation $g^2/(4\pi)=\alpha_s$. 
Further we shall denote $L_2(F)$ the set of the modulo square integrable
complex functions on any manifold $F$ furnished with an integration measure, 
then $L^n_2(F)$ will be the $n$-fold direct product of $L_2(F)$
endowed with the obvious scalar product while $\dag$ and $\ast$ stand, 
respectively, for Hermitian and complex conjugation. Our choice of Dirac 
$\gamma$-matrices conforms to the so-called standard representation and is 
$$\gamma^0=\pmatrix{1&0\cr 0&-1\cr}\,,
\gamma^b=\pmatrix{0&\sigma_b\cr-\sigma_b&0\cr}\,,
b= 1,2,3\>, {\bf \alpha}=\gamma^0{\bf\gamma}=
\pmatrix{0 & {\bf\sigma}\cr{\bf\sigma}&0\cr}\,,
\eqno(2)$$
where $\sigma_b$ denote the ordinary Pauli matrices and ${\bf \sigma}=
\sigma_1{\bf i}+\sigma_2{\bf j}+\sigma_3{\bf k}$. At last $\otimes$ means 
tensorial product of matrices and $I_3$ is the unit $3\times3$ matrix. 

When calculating we apply the 
relations $1\ {\rm GeV^{-1}}\approx0.1973269679\ {\rm fm}\>$,
$1\ {\rm s^{-1}}\approx0.658211915\times10^{-24}\ {\rm GeV}\>$, 
$1\ {\rm V/m}\approx0.2309956375\times 10^{-23}\ {\rm GeV}^2$, 
$1\ {\rm T}\approx0.6925075988\times 10^{-15}\ {\rm GeV}^2$.

Finally, for the necessary estimates we shall employ the $T_{00}$-component 
(volumetric energy density ) of the energy-momentum tensor for a 
SU(3)-Yang-Mills field which should be written in the chosen system of units 
in the form
$$T_{\mu\nu}=-F^a_{\mu\alpha}\,F^a_{\nu\beta}\,g^{\alpha\beta}+
{1\over4}F^a_{\beta\gamma}\,F^a_{\alpha\delta}g^{\alpha\beta}g^{\gamma\delta}
g_{\mu\nu}\>. \eqno(3) $$

\section{Main relations}

As was mentioned above, our considerations shall be based on the unique family 
of compatible nonperturbative solutions for 
the Dirac-Yang-Mills system (derived from QCD-Lagrangian) studied in details 
in Refs. \cite{{Gon01},{Gon051},{Gon052}}.  Referring for more details to those 
references, let us briefly decribe only the relations necessary to us in 
the presemt Letter following the notations from \cite{{Gon052}}. 

One part of the mentioned family is presented by the unique nonperturbative 
confining solution of the Yang-Mills equations and looks as follows 
$$ A^3_t+\frac{1}{\sqrt{3}}A^8_t =-\frac{a_1}{r}+A_1 \>,
 -A^3_t+\frac{1}{\sqrt{3}}A^8_t=-\frac{a_2}{r}+A_2\>,
-\frac{2}{\sqrt{3}}A^8_t=\frac{a_1+a_2}{r}-(A_1+A_2)\>, $$
$$ A^3_\varphi+\frac{1}{\sqrt{3}}A^8_\varphi =b_1r+B_1 \>,
 -A^3_\varphi+\frac{1}{\sqrt{3}}A^8_\varphi=b_2r+B_2\>,
-\frac{2}{\sqrt{3}}A^8_\varphi=-(b_1+b_2)r-(B_1+B_2)\> \eqno(4)$$
with the real constants $a_j, A_j, b_j, B_j$ parametrizing the family. 

Another part of the family is given by the unique nonperturbative modulo 
square integrable solutions of the Dirac equation in the confining 
SU(3)-field of (4) $\Psi=(\Psi_1, \Psi_2, \Psi_3)$ 
with the four-dimensional Dirac spinors 
$\Psi_j$ representing the $j$th colour component of the meson, 
which may describe the relativistic bound states of two quarks (mesons) 
and look as follows (with Pauli matrix $\sigma_1$)
$$\Psi_j=e^{i\omega_j t}r^{-1}\pmatrix{F_{j1}(r)\Phi_j(\vartheta,\varphi)\cr\
F_{j2}(r)\sigma_1\Phi_j(\vartheta,\varphi)}\>,j=1,2,3\eqno(5)$$
with the 2D eigenspinor $\Phi_j=\pmatrix{\Phi_{j1}\cr\Phi_{j2}}$ of the
euclidean Dirac operator ${\cal D}_0$ on the unit sphere ${\Bbb S}^2$, while 
the coordinate $r$ stands for the distance between quarks.
The explicit form of $\Phi_j$ is not needed here and
can be found in Ref. \cite{{Gon052},{Gon99}}. For the purpose of the present 
Letter we shall adduce the necessary spinors below. Spinors $\Phi_j$ form an 
orthonormal basis in $L_2^2({\Bbb S}^2)$. 

The energy spectrum of a meson is given 
by $\omega=\omega_1+\omega_2+\omega_3$ with 
$$\omega_j=\omega_j(n_j,l_j,\lambda_j)=gA_j+$$ 
$$\frac{-\Lambda_j g^2a_jb_j\pm(n_j+\alpha_j)
\sqrt{(n_j^2+2n_j\alpha_j+\Lambda_j^2)\mu_0^2+g^2b_j^2(n_j^2+2n_j\alpha_j)}}
{n_j^2+2n_j\alpha_j+\Lambda_j^2}\>, j=1,2,3\>,\eqno(6)$$

where $g$ is the gauge coupling constant, $a_3=-(a_1+a_2)$, $b_3=-(b_1+b_2)$, 
$A_3=-(A_1+A_2)$, $B_3=-(B_1+B_2)$, 
$\Lambda_j=\lambda_j-gB_j$, $\alpha_j=\sqrt{\Lambda_j^2-g^2a_j^2}$, 
$n_j=0,1,2,...$, while $\lambda_j=\pm(l_j+1)$ are
the eigenvalues of euclidean Dirac operator ${\cal D}_0$ 
on unit sphere with $l_j=0,1,2,...$, $\mu_0$ is a mass parameter and one can 
consider it to be the reduced
mass which is equal, {\it e. g.} for quarkonia, to half the 
current mass of quarks forming a quarkonium.  
As is clear from (6), parameters $A_{1,2}$ of solution (4) only
shift the origin of count for the corresponding energies and we can 
consider $A_1=A_2=0$.

Within the given Letter we need only the radial parts of (5) at $n_j=0$ that 
are
$$F_{j1}=C_jP_jr^{\alpha_j}e^{-\beta_jr}\left(1-
\frac{gb_j}{\beta_j}\right), P_j=gb_j+\beta_j, $$
$$F_{j2}=iC_jQ_jr^{\alpha_j}e^{-\beta_jr}\left(1+
\frac{gb_j}{\beta_j}\right), Q_j=\mu_0+\omega_j\eqno(7)$$
with $\beta_j=\sqrt{\mu_0^2-\omega_j^2+g^2b_j^2}$ while $C_j$ is determined 
from the normalization condition
$\int_0^\infty(|F_{j1}|^2+|F_{j2}|^2)dr=\frac{1}{3}$. 
Consequently, we shall gain that
$\Psi_j\in L_2^{4}({\Bbb R}^3)$ at any $t\in{\Bbb R}$ and, as a result,
the solutions of (5) may describe relativistic bound states (mesons) 
with the energy (mass) spectrum (6).

It is useful to obtain the nonrelativistic limit (when 
$c\to\infty$) for spectrum (6). For that one should replace 
$g\to g/\sqrt{\hbar c}$, 
$a_j\to a_j/\sqrt{\hbar c}$, $b_j\to b_j\sqrt{\hbar c}$, 
$B_j\to B_j/\sqrt{\hbar c}$ and, expanding (6) in $z=1/c$, we shall get
$$\omega_j(n_j,l_j,\lambda_j)=$$
$$\pm\mu_0c^2\left[1\mp
\frac{g^2a_j^2}{2\hbar^2(n_j+|\lambda_j|)^2}z^2\right]
-\left[\frac{\lambda g^2a_jb_j}{\hbar(n_j+|\lambda_j|)^2}\,
\pm\mu_0\frac{g^3B_ja_j^2f(n_j,\lambda_j)}{\hbar^3(n_j+|\lambda_j|)^{7}}\right]
z\,+O(z^2)\>,\eqno(8)$$
where 
$f(n_j,\lambda_j)=4\lambda_jn_j(n_j^2+\lambda_j^2)+
\frac{|\lambda_j|}{\lambda_j}\left(n_j^{4}+6n_j^2\lambda_j^2+\lambda_j^4
\right)$. 

As is seen from (8), at $c\to\infty$ the contribution of linear magnetic 
colour field (parameters $b_j, B_j$) to spectrum really vanishes and spectrum 
in essence becomes purely Coulomb one (modulo the rest energy). Also it is 
clear that when $n_j\to\infty$, $\omega_j\to\pm\sqrt{\mu_0^2+g^2b_j^2}$. 

We may seemingly use (6) with various combinations of signes ($\pm$) before 
second summand in numerators of (6) but, due to (8), it is 
reasonable to take all signs equal to $+$ which is our choice within the 
Letter. Besides, 
as is not complicated to see, radial parts in nonrelativistic limit have 
the behaviour of form $F_{j1},F_{j2}\sim r^{l_j}$, which allows one to call 
quantum number $l_j$ angular momentum for $j$th colour component though angular 
momentum is not conserved in the field (4) \cite{{Gon01},{Gon052}}. So for 
mesons under consideration we should put all $l_j=0$. 

Finally it should be noted that spectrum (6) is degenerated owing to 
degeneracy of eigenvalues for the
euclidean Dirac operator ${\cal D}_0$ on the unit sphere ${\Bbb S}^2$. Namely,  
each eigenvlalue of ${\cal D}_0$ $\lambda =\pm(l+1), l=0,1,2...$, has 
multiplicity $2(l+1)$ so we has $2(l+1)$ eigenspinors orthogonal to each other. 
Ad referendum we need eigenspinors corresponding to $\lambda =\pm1$ ($l=0$) 
so here is their explicit form 
$$\lambda=-1: \Phi=\frac{C}{2}\pmatrix{e^{i\frac{\vartheta}{2}}
\cr e^{-i\frac{\vartheta}{2}}\cr}e^{i\varphi/2},\> {\rm or}\>\>
\Phi=\frac{C}{2}\pmatrix{e^{i\frac{\vartheta}{2}}\cr
-e^{-i\frac{\vartheta}{2}}\cr}e^{-i\varphi/2},$$
$$\lambda=1: \Phi=\frac{C}{2}\pmatrix{e^{-i\frac{\vartheta}{2}}\cr
e^{i\frac{\vartheta}{2}}\cr}e^{i\varphi/2}, \> {\rm or}\>\>
\Phi=\frac{C}{2}\pmatrix{-e^{-i\frac{\vartheta}{2}}\cr
e^{i\frac{\vartheta}{2}}\cr}e^{-i\varphi/2} 
\eqno(9) $$
with the coefficient $C=1/\sqrt{2\pi}$ (for more details see 
Refs. \cite{{Gon052},{Gon99}}).

\section{Estimates for parameters of SU(3)-gluonic field}

Within the present Letter we shall use relations (6) at $n_j=0=l_j$ so energy 
(mass) of mesons under consideration is given by 
 
$$\mu=\sum\limits_{j=1}^3\omega_j(0,0,\lambda_j)=\sum\limits_{j=1}^3
\left(\frac{-g^2a_jb_j}{\Lambda_j}+\frac{\alpha_j\mu_0}{|\Lambda_j|}\right)
\>\eqno(10)$$
and, as a consequence, the corresponding meson wave functions of (5) are 
represented by (7), (9). 
\subsection{Choice of quark masses and gauge coupling constant}
It is evident for employing the above relations we have to assign some values 
to quark masses and gauge coupling constant $g$. In accordance with 
Ref. \cite{pdg}, at present the current quark masses are restricted 
to intervals $1.5\>{\rm MeV}\le m_u\le 5\> {\rm MeV}$, 
$3.0\> {\rm MeV}\le m_d\le 9 \>{\rm MeV}$, 
$60 \>{\rm MeV}\le m_s\le 170 \>{\rm MeV}$, 
so we take $m_u=(1.5+5)/2\> {\rm MeV}=3.25\>{\rm MeV}$, 
$m_d=(3+9)/2\> {\rm MeV}=6\>{\rm MeV}$, $m_s=(60+170)/2\> {\rm MeV}=
115 \>{\rm MeV}$. 
Under the circumstances, the reduced mass $\mu_0$ of Table 1 will respectively 
take values 
$m_u/2, m_d/2, m_um_d/(m_u+m_d), m_um_s/(m_u+m_s),m_dm_s/(m_d+m_s)$. As to 
gauge coupling constant $g=\sqrt{4\pi\alpha_s}$, at present little is known about 
the values of the strong coupling constant $\alpha_s=\alpha_s(Q^2)$ at the 
momentum transfer $Q\to 0$. We use the results of recent works \cite{De} on 
extracting $\alpha_s$ from the so-called Bjorken sum rule, wherefrom one can 
conclude that $\alpha_s\to \pi=3.1415...$ when $Q\to 0$, i. e., 
$g\to{2\pi}=6.2831...$. An extrapolation of the 
results of \cite{De} to the mass scales of mesons under discussion gives rise 
to the values of $g$ adduced in Table 1.
\subsection{Electric formfactor}
For each meson with the wave function $\Psi=(\Psi_j)$ of (5) we can define 
electromagnetic current $J^\mu=\overline{\Psi}(\gamma^\mu\otimes I_3)\Psi=
(\Psi^{\dag}\Psi,\Psi^{\dag}({\bf \alpha}\otimes I_3)\Psi)=(\rho,{\bf J})$. 
Electric formfactor $f(K)$ is the Fourier transform of $\rho$
$$ f(K)= \int\Psi^{\dag}\Psi e^{-i{\bf K}{\bf r}}d^3x=\sum\limits_{j=1}^3
\int\Psi_j^{\dag}\Psi_j e^{-i{\bf K}{\bf r}}d^3x =\sum\limits_{j=1}^3f_j(K)=$$ 
$$\sum\limits_{j=1}^3
\int (|F_{j1}|^2+|F_{j2}|^2)\Phi_j^{\dag}\Phi_j
\frac{e^{-i{\bf K}{\bf r}}}{r^2}d^3x,\>
d^3x=r^2\sin{\vartheta}dr d\vartheta d\varphi\eqno(11)$$
with the momentum transfer $K$. We can consider vector ${\bf K}$ to be 
directed along z-axis. Then ${\bf Kr}=Kr\cos{\vartheta}$ and 
at $n_j=0=l_j$, as is easily seen, for any  
spinor of (9) we have $\Phi_j^{\dag}\Phi_j=1/(4\pi)$, so with the help of 
(7) and relations (see Ref. \cite{PBM1}): $\int_0^\infty 
r^{\alpha-1}e^{-pr}dr=
\Gamma(\alpha)p^{-\alpha}$, Re $\alpha,p >0$, 
$\int_0^\infty r^{\alpha-1}e^{-pr}\sin{(Kr)}dr=
\Gamma(\alpha)(K^2+p^2)^{-\alpha/2}\sin{\arctan{(K/p)}}$, Re $\alpha >-1$, 
Re $p > |{\rm Im}\, K|$, $\Gamma(\alpha+1)=\alpha\Gamma(\alpha)$, 
$\int_0^\pi e^{-iKr\cos{\vartheta}}\sin{\vartheta}d\vartheta=2\sin{(Kr)}/(Kr)$, 
we shall obtain 
$$ f(K)=\sum\limits_{j=1}^3f_j(K)=
\sum\limits_{j=1}^3\frac{(2\beta_j)^{2\alpha_j+1}}{6\alpha_j}\cdot
\frac{\sin{[2\alpha_j\arctan{(K/(2\beta_j))]}}}{K(K^2+4\beta_j^2)^{\alpha_j}}$$
$$=\sum\limits_{j=1}^3\left(\frac{1}{3}-\frac{2\alpha^2_j+3\alpha_j+1}
{6\beta_j^2}\cdot \frac{K^2}{6}\right)+O(K^4), \eqno(12)$$
wherefrom it is clear that $f(K)$ is a function of $K^2$, as should be, and 
we can determine the root-mean-square radius of meson in the form 
$$<r>=\sqrt{\sum\limits_{j=1}^3\frac{2\alpha^2_j+3\alpha_j+1}
{6\beta_j^2}}.\eqno(13)$$
Of course, we can directly calculate $<r>$ in accordance with the standard 
quantum mechanics rules as $<r>=\sqrt{\int r^2\Psi^{\dag}\Psi d^3x}=
\sqrt{\sum\limits_{j=1}^3\int r^2\Psi^{\dag}_j\Psi_j d^3x}$ and the 
result will be the same as in (13). Therefore, we should not call $<r>$ of (13) 
the {\em charge} radius of meson -- it is just the radius of meson determined 
by the wave functions of (5) (at $n_j=0=l_j$) with respect to strong interaction. 

\subsection{Magnetic moment}
We can define the volumetric magnetic moment density by 
${\bf m}=q({\bf r}\times {\bf J})/2=q[(yJ_z-zJ_y){\bf i}+
(zJ_x-xJ_z){\bf j}+(xJ_y-yJ_x){\bf k}]/2$ with the meson charge $q$ and 
${\bf J}=\Psi^{\dag}({\bf \alpha}\otimes I_3)\Psi$. Using (5) we have in the 
explicit form 
$$J_x=\sum\limits_{j=1}^3
\int (F^\ast_{j1}F_{j2}+F^\ast_{j2}F_{j1})\frac{\Phi_j^{\dag}\Phi_j}
{r^2}d^3x,\> 
J_y=\sum\limits_{j=1}^3
\int (F^\ast_{j1}F_{j2}-F^\ast_{j2}F_{j1})
\frac{\Phi_j^{\dag}\sigma_2\sigma_1\Phi_j}{r^2}d^3x,\>$$
$$J_z=\sum\limits_{j=1}^3
\int (F^\ast_{j1}F_{j2}-F^\ast_{j2}F_{j1})
\frac{\Phi_j^{\dag}\sigma_3\sigma_1\Phi_j}{r^2}d^3x.  \eqno(14)$$
Magnetic moment of meson is ${\bf M}=\int_V {\bf m}d^3x$, where $V$ is volume 
of meson. Then at $l_j=0$ we have $J_y=0$ for any  
spinor of (9), while $\int_V m_{x,y,z}d^3x=0$ because of 
turning to zero either integral over $\vartheta$ or the one over $\varphi$, 
which is easily to check. 
As a result, magnetic moments of mesons under consideration with the 
wave functions of (5) (at $l_j=0$) are equal to zero, as should be according 
to experimental data \cite{pdg}. 
\subsection{Numerical results}
We employed relations (10) and (13) for obtaining estimates of the confining 
SU(3)-gluonic field parameters in mesons under discussion and, to impose more 
restrictions, we considered in (10) each $\omega_j=\mu/3$, $\mu$ is meson mass, 
though it is not obligatory. Also the experimental estimates, if any, of $<r>$ 
were used from Refs. \cite{{pdg},{Pi}}. The results are adduced in Tables 1, 2. 
It should be noted that in accordance with SQM \cite{RF} 
$\pi^0$ =($\overline{u}u-\overline{d}d)/\sqrt{2}$ is equiprobable superposition 
of two quarkonia, consequently, we have two set of parameters and, to reach 
orthogonality for $\overline{u}u$ and $\overline{d}d$ states, we should assign 
different spinors of (9) at $\lambda=-1$ to the corresponding wave functions 
of $\pi^0$.  Besides, according to SQM the all $\pi$-meson states are orthogonal 
to each other and, to reach it, we should assign 
different spinors of (9) at $\lambda=1$ to the corresponding wave functions 
of $\pi^\pm$. Analogous situation is for $K$-meson sector. 

\begin{table}[htbp]
\caption{Gauge coupling constant, mass parameter $\mu_0$ and
parameters of the confining SU(3)-gluonic field for pions and kaons.}
\label{t.1}
\begin{center}
\begin{tabular}{|c|c|c|c|c|c|c|c|c|}
\hline
\small Particle & \small $ g$ & \small $\mu_0$ (\small MeV) & \small $a_1$ 
& \small $a_2$ & \small $b_1 (\small GeV)$ 
& \small $b_2$ (\small GeV) & \small $B_1$ & \small $B_2$ \\
\hline
$\pi^{0}$---$\overline{u}u$  & \scriptsize 6.2816 & \scriptsize 1.625 & 
\scriptsize -0.0124758
& \scriptsize -0.00630731 & \scriptsize -0.275416 & \scriptsize 0.211701 & 
\scriptsize 0.3385
& \scriptsize  -0.3526 \\
\hline
$\pi^{0}$---$\overline{d}d$  & \scriptsize 6.2816 & \scriptsize 3.00 & 
\scriptsize 0.0121524
& \scriptsize 0.00614414 & \scriptsize 0.273783 & \scriptsize -0.210436 & 
\scriptsize 0.3385
& \scriptsize  -0.3526 \\
\hline
$\pi^{\pm}$---$u\overline{d}$, $\overline{u}d$ 
  & \scriptsize 6.2745 & \scriptsize 2.10811 & \scriptsize 0.00651124 
& \scriptsize 0.0128925 & \scriptsize 0.21007 & \scriptsize -0.273365 & 
\scriptsize 0.3526
& \scriptsize  -0.3385 \\
\hline
$K^{\pm}$---$u\overline{s}$, $\overline{u}s$ 
  & \scriptsize 6.1256 & \scriptsize 3.16068 & \scriptsize 0.0348355
& \scriptsize 0.0523979 & \scriptsize 0.524585 & \scriptsize -0.366167 & 
\scriptsize 0.5303 & \scriptsize  -0.8914 \\
\hline
$K^{0}$, $\overline{K}^{0}$---$d\overline{s}$, $\overline{d}s$ 
  & \scriptsize 6.11497 & \scriptsize 5.70248 & \scriptsize 0.0494856
& \scriptsize 0.0240874 & \scriptsize 0.104398 & \scriptsize -0.263159 & 
\scriptsize 0.36052 & \scriptsize  -0.784 \\
\hline
\end{tabular}
\end{center}
\end{table}

\begin{table}[htbp]
\caption{Theoretical and experimental meson masses and radii}
\label{t.2}
\begin{center}
\begin{tabular}{|c|c|c|c|c|} 
\hline
\tiny Particle & \tiny Theoret. (MeV) &  \tiny Experim. (MeV) & 
\tiny Theoret. $<r>$ (fm)  & \tiny Experim. $<r>$ (fm)\\
\hline
\scriptsize $\pi^{0}$---$\overline{u}u$   & \scriptsize $\mu= \omega_1(0,0,-1)+
\omega_2(0,0,-1)+
\omega_3(0,0,-1)= 134.976$ & \scriptsize 134.976 & \scriptsize 0.602594 & 
\scriptsize -- \\
\hline
\scriptsize $\pi^{0}$---$\overline{d}d$ & \scriptsize $\mu =\omega_1(0,0,-1)+
\omega_2(0,0,-1)+
\omega_3(0,0,-1)= 134.976$ & \scriptsize 134.976 & \scriptsize 0.606236 & 
\scriptsize --\\                                           
\hline
\scriptsize $\pi^{\pm}$---$u\overline{d}$, $\overline{u}d$  & 
\scriptsize $\mu =\omega_1(0,0,1)+
\omega_2(0,0,1)+\omega_3(0,0,1)= 139.570 $ & \scriptsize 139.56995 & 
\scriptsize 0.607418 & \scriptsize 0.6050\\ 
\hline
\scriptsize $K^{\pm}$---$u\overline{s}$, $\overline{u}s$  & \scriptsize $\mu =
\omega_1(0,0,-1)+\omega_2(0,0,-1)+\omega_3(0,0,-1)= 493.677 $ & \scriptsize 
493.677 & \scriptsize 0.564046 & \scriptsize 0.560\\
\hline
\scriptsize $K^{0}$, $\overline{K}^{0}$---$d\overline{s}$, $\overline{d}s$  & 
\scriptsize $\mu =\omega_1(0,0,1)+\omega_2(0,0,1)+\omega_3(0,0,1)= 497.672 $ 
& \scriptsize 497.672 & \scriptsize 0.560964 & --\\
\hline
\end{tabular}
\end{center}
\end{table}

\section{Estimates of gluon concentrations, electric and magnetic colour field 
strengths}
To obtain further physical characteristics of the confining SU(3)-gluonic 
field in pions and kaons let us remind that, according to Refs. 
\cite{{GB04},{Gon051},{Gon052}}, we can confront the field (4) with the 
3-dimensional SU(3)-Lie algebra valued 1-forms of electric {\bf E} and magnetic 
{\bf H} colour fields and also with $T_{00}$-component (volumetric energy 
density) of the energy-momentum tensor (3)

$$ {\bf E}=\left[\lambda_3(a_1-a_2)+
\lambda_8(a_1+a_2)\sqrt{3}\right]\frac{dr}{2r^2},\> 
{\bf H}=-\left[\lambda_3(b_1-b_2)+
\lambda_8(b_1+b_2)\sqrt{3}\right]
\frac{d\vartheta}{2\sin{\vartheta}}, 
\eqno(15)$$
$$T_{00}\equiv T_{tt}=
\frac{1}{2}\left(\frac{a_1^2+a_1a_2+a_2^2}{r^4}+
\frac{b_1^2+b_1b_2+b_2^2}{r^2\sin^2{\vartheta}}\right)\equiv
\frac{{\cal A}}{r^4}+
\frac{{\cal B}}{r^2\sin^2{\vartheta}}\>\eqno(16)$$
with real ${\cal A}>0$, ${\cal B}>0$. When defining the scalar product for the 
SU(3)-Lie algebra valued 1-forms $A=A^a_\mu\lambda_adx^\mu$, 
$B=B^b_\nu\lambda_bdx^\nu$ by $G(A,B)=
g^{\mu\nu}A^a_\mu B^b_\nu{\rm Tr}(\lambda_a\lambda_b)/2=
g^{\mu\nu}A^a_\mu B^b_\nu\delta_{ab}$, we shall have electric and magnetic 
colour field strenghts modulo equal to
$$E=\sqrt{G({\bf E},{\bf E})}=\frac{\sqrt{a_1^2+a_1a_2+a_2^2}}{r^2},\>
H=\sqrt{G({\bf H},{\bf H})}=
\frac{\sqrt{b_1^2+b_1b_2+b_2^2}}{r\sin{\vartheta}}, \eqno(17)$$
so that $T_{00}=[G({\bf E},{\bf E})+G({\bf H},{\bf H})]/2=(E^2+H^2)/2$. 

To estimate the gluon concentrations
we can employ $T_{00}$-component of (16) and, taking the quantity
$\omega= \mit\Gamma$, the whole decay width of a meson, for 
the characteristic frequency of gluons we obtain
the sought characteristic concentration $n$ in the form
$$n=\frac{T_{00}}{\mit\Gamma}\> \eqno(18)$$
so we can rewrite
(16) in the form
$T_{00}=T_{00}^{\rm coul}+T_{00}^{\rm lin}$ conforming to the contributions 
from the Coulomb and linear parts of the
solution (4). The latter gives the corresponding split of $n$ from (18) as 
$n=n_{\rm coul} + n_{\rm lin}$. 

The parameters of Tables 1, 2 were employed when computing and for simplicity 
we put $\sin{\vartheta}=1$ in (16)--(17). Also there were used the following 
present-day whole decay widths of mesons under 
consideration \cite{pdg}: ${\mit\Gamma}=1/\tau$ with the life times $\tau= 
8.4\times10^{-17}$ s, 2.6033$\times10^{-8}$ s, 1.2386$\times10^{-8}$ s, 
0.8953$\times10^{-10}$ s ($K^0_S$-mode), 5.18$\times10^{-8}$ s ($K^0_L$-mode), 
respectively, whereas the Bohr radius 
$a_0=0.529177249\cdot10^{5}\ {\rm fm}$ \cite{pdg}. At last, as has been 
discussed in Refs. \cite{{GB04},{Gon052}}, 
we can estimate the quark velocities in the mesons under exploration from the 
condition 
$$v_q=\frac{1}{\sqrt{1+\left(\frac{\lambda_B}{\lambda_q}\right)^2}}\>
\eqno(19)$$
with the quark Compton wavelength 
$\lambda_q=1/m_q$ while we take the quark de Broglie 
wavelength $\lambda_B=0.1r_0$ with $r_0=<r>$ from Table 2. 

Tables 3, 4 contain the numerical results for $n_{\rm coul}$, $n_{\rm lin}$, $n$, 
$E$, $H$, $v_q$ for the mesons under discussion.
\begin{table}[htbp]
\caption{Gluon concentrations, electric and magnetic colour field strengths in 
pions.}
\label{t.3}
\begin{center}
\begin{tabular}{|llllll|}
\hline
\scriptsize $\pi^{0}$---$\overline{u}u$: & \scriptsize 
$r_0=<r>= 0.602594\ {\rm fm}$, &  
\scriptsize $v_u = 0.999504$ & &  &\\
\hline 
\tiny $r$ & \tiny $n_{\rm coul}$ & \tiny $n_{\rm lin}$ 
& \tiny $n$ & \tiny $E$ & \tiny $H$ \\
\tiny (fm) & \tiny $ ({\rm m}^{-3}) $ 
& \tiny (${\rm m}^{-3}) $ & \tiny (${\rm m}^{-3}) $ 
& \tiny $({\rm V/m})$ & \tiny $({\rm T})$\\
\hline
\tiny $0.1r_0$ & \tiny $ 0.172647\times10^{57}$   
& \tiny $ 0.142633\times10^{57}$ & \tiny $ 0.315280\times10^{57}$ 
& \tiny $ 0.768576\times10^{23}$  & \tiny $ 0.118089\times10^{16}$ \\
\hline
\tiny$r_0$ & \tiny$ 0.172647\times10^{53}$ & \tiny$ 0.142633\times10^{55}$ 
& \tiny$ 0.144359\times10^{55}$& \tiny$0.768576\times10^{21}$  
& \tiny$ 0.118089\times10^{15}$  \\
\hline
\tiny$1.0$ & \tiny$ 0.227646\times10^{52}$  & \tiny$ 0.517927\times10^{54}$ 
& \tiny$0.520203\times10^{54}$ & \tiny$0.279085\times10^{21}$  
& \tiny$0.711597\times10^{14}$  \\
\hline
\tiny$10r_0$ & \tiny$0.172647\times10^{49}$  
& \tiny$0.142633\times10^{53}$ & \tiny$0.142650\times10^{53}$ 
& \tiny$0.768576\times10^{19}$  & \tiny$0.118089\times10^{14}$  \\
\hline
\tiny$a_0$ & \tiny$0.290305\times10^{33}$  & \tiny$0.184955\times10^{45}$ & 
\tiny$0.184955\times10^{45}$ & \tiny$0.996631\times10^{12}$ 
& \tiny$0.134472\times10^{10}$  \\
\hline
\hline
\scriptsize $\pi^{0}$---$\overline{d}d$: & \scriptsize 
$r_0=<r>= 0.606236\ {\rm fm}$,  &  
\scriptsize $v_d = 0.999080 $ & &  &\\
\hline 
\tiny $r$ & \tiny $n_{\rm coul}$ & \tiny $n_{\rm lin}$ 
& \tiny $n$ & \tiny $E$ & \tiny $H$ \\
\tiny (fm) & \tiny $ ({\rm m}^{-3}) $ 
& \tiny (${\rm m}^{-3}) $ & \tiny (${\rm m}^{-3}) $ 
& \tiny $({\rm V/m})$ & \tiny $({\rm T})$\\
\hline
\tiny $0.1r_0$ & \tiny $0.159916\times10^{57}$   
& \tiny $0.139255\times10^{57}$ & \tiny$0.299171\times10^{57}$ 
& \tiny $0.739697\times10^{23}$  & \tiny $0.116682\times10^{16}$ \\
\hline
\tiny$r_0$ & \tiny$0.159916\times10^{53}$ & \tiny$0.139255\times10^{55}$ 
& \tiny$0.140854\times10^{55}$& \tiny$0.739697\times10^{21}$  
& \tiny$0.116682\times10^{15}$  \\
\hline
\tiny$1.0$ & \tiny$0.216003\times10^{52}$  & \tiny$0.511791\times10^{54}$ 
& \tiny$0.513951\times10^{54}$ & \tiny$0.271855\times10^{21}$  
& \tiny$0.707369\times10^{14}$  \\
\hline
\tiny$10r_0$ & \tiny$0.159916\times10^{49}$  
& \tiny$0.139255\times10^{53}$ & \tiny$0.139270\times10^{53}$ 
& \tiny$0.739697\times10^{19}$  & \tiny$0.116682\times10^{14}$  \\
\hline
\tiny$a_0$ & \tiny$0.275458\times10^{33}$  & \tiny$0.182764\times10^{45}$ & 
\tiny$0.182764\times10^{45}$ & \tiny$0.970811\times10^{11}$ 
& \tiny$0.133673\times10^{10}$  \\
\hline
\hline
\scriptsize $\pi^{\pm}$---$u\overline{d}$, $\overline{u}d$: & \scriptsize 
$r_0=<r>= 0.607418\ {\rm fm}$,  & \scriptsize $v_u = 0.999500$,&  
\scriptsize $v_d= 0.999078 $  &  &\\
\hline 
\tiny $r$ & \tiny $n_{\rm coul}$ & \tiny $n_{\rm lin}$ 
& \tiny $n$ & \tiny $E$ & \tiny $H$ \\
\tiny (fm) & \tiny $ ({\rm m}^{-3}) $ 
& \tiny (${\rm m}^{-3}) $ & \tiny (${\rm m}^{-3}) $ 
& \tiny $({\rm V/m})$ & \tiny $({\rm T})$\\
\hline
\tiny $0.1r_0$ & \tiny $0.553136\times10^{65}$   
& \tiny $0.428537\times10^{65}$ & \tiny$0.981673\times10^{65}$ 
& \tiny $0.781449\times10^{23}$  & \tiny $0.116271\times10^{16}$ \\
\hline
\tiny$r_0$ & \tiny$0.553136\times10^{61}$ & \tiny$0.428537\times10^{63}$ 
& \tiny$0.434069\times10^{63}$& \tiny$0.781449\times10^{21}$  
& \tiny$0.116271\times10^{15}$  \\
\hline
\tiny$1.0$ & \tiny$0.752978\times10^{60}$  & \tiny$0.158112\times10^{63}$ 
& \tiny$0.158865\times10^{63}$ & \tiny$0.288321\times10^{21}$  
& \tiny$0.706251\times10^{14}$  \\
\hline
\tiny$10r_0$ & \tiny$0.553136\times10^{57}$  
& \tiny$0.428537\times10^{61}$ & \tiny$0.428593\times10^{61}$ 
& \tiny$0.781449\times10^{19}$  & \tiny$0.116271\times10^{14}$  \\
\hline
\tiny$a_0$ & \tiny$0.960236\times10^{41}$  & \tiny$0.564627\times10^{53}$ & 
\tiny$0.564627\times10^{53}$ & \tiny$0.102961\times10^{12}$ 
& \tiny$0.133462\times10^{10}$  \\
\hline
\end{tabular}
\end{center}
\end{table}

\begin{table}[htbp]
\caption{Gluon concentrations, electric and magnetic colour field strengths in 
kaons.}
\label{t.4}
\begin{center}
\begin{tabular}{|llllll|}
\hline
\scriptsize $K^{\pm}$---$u\overline{s}$, $\overline{u}s$: & \scriptsize 
$r_0=<r>= 0.564046 \ {\rm fm}$, & \scriptsize $v_u= 0.999536 $, & 
\scriptsize $v_s= 0.983958 $ &  &  \\ 
\hline 
\tiny $r$ & \tiny $n_{\rm coul}$ & \tiny $n_{\rm lin}$ 
& \tiny $n$ & \tiny $E$ & \tiny $H$ \\
\tiny (fm) & \tiny $ ({\rm m}^{-3}) $ 
& \tiny (${\rm m}^{-3}) $ & \tiny (${\rm m}^{-3}) $ 
& \tiny $({\rm V/m})$ & \tiny $({\rm T})$\\
\hline
\tiny $0.1r_0$ & \tiny $0.699798\times10^{66}$   
& \tiny $0.835931\times10^{65}$ & \tiny$0.783392\times10^{66}$ 
& \tiny $0.402965\times10^{24}$  & \tiny $0.235429\times10^{16}$ \\
\hline
\tiny$r_0$ & \tiny$0.699799\times10^{62}$ & \tiny$0.835931\times10^{63}$ 
& \tiny$0.905911\times10^{63}$& \tiny$0.402965\times10^{22}$  
& \tiny$0.235429\times10^{15}$  \\
\hline
\tiny$1.0$ & \tiny$0.708323\times10^{61}$  & \tiny$0.265950\times10^{63}$ 
& \tiny$0.273033\times10^{63}$ & \tiny$0.128203\times10^{22}$  
& \tiny$0.132793\times10^{15}$  \\
\hline
\tiny$10r_0$ & \tiny$0.699799\times10^{58}$  
& \tiny$0.835931\times10^{61}$ & \tiny$0.836631\times10^{61}$ 
& \tiny$0.402965\times10^{20}$  & \tiny$0.235429\times10^{14}$  \\
\hline
\tiny$a_0$ & \tiny$0.903288\times10^{42}$  & \tiny$0.949724\times10^{53}$ & 
\tiny$0.949724\times10^{53}$ & \tiny$0.457820\times10^{12}$ 
& \tiny$0.250942\times10^{10}$  \\
\hline
\hline
\scriptsize $K^0$, $\overline{K}^{0}$---$d\overline{s}$, $\overline{d}s$ 
($K^0_S$--mode): & \scriptsize 
$r_0=<r>= 0.560964\ {\rm fm}$,  & \scriptsize $v_d= 0.999148 $,& 
\scriptsize $v_s= 0.984044 $  &  &\\
\hline 
\tiny $r$ & \tiny $n_{\rm coul}$ & \tiny $n_{\rm lin}$ 
& \tiny $n$ & \tiny $E$ & \tiny $H$ \\
\tiny (fm) & \tiny $ ({\rm m}^{-3}) $ 
& \tiny (${\rm m}^{-3}) $ & \tiny (${\rm m}^{-3}) $ 
& \tiny $({\rm V/m})$ & \tiny $({\rm T})$\\
\hline
\tiny $0.1r_0$ & \tiny $0.377302\times10^{64}$   
& \tiny $0.148175\times10^{63}$ & \tiny$0.392120\times10^{64}$ 
& \tiny $0.348022\times10^{24}$  & \tiny $0.116585\times10^{16}$ \\
\hline
\tiny$r_0$ & \tiny$0.377302\times10^{60}$ & \tiny$0.148175\times10^{61}$ 
& \tiny$0.185905\times10^{61}$& \tiny$0.348022\times10^{22}$  
& \tiny$0.116585\times10^{15}$  \\
\hline
\tiny$1.0$ & \tiny$0.373620\times10^{59}$  & \tiny$0.466279\times10^{60}$ 
& \tiny$0.503641\times10^{60}$ & \tiny$0.109516\times10^{22}$  
& \tiny$0.654000\times10^{14}$  \\
\hline
\tiny$10r_0$ & \tiny$0.377302\times10^{56}$  
& \tiny$0.148175\times10^{59}$ & \tiny$0.148553\times10^{59}$ 
& \tiny$0.348022\times10^{20}$  & \tiny$0.116585\times10^{14}$  \\
\hline
\tiny$a_0$ & \tiny$0.476458\times10^{40}$  & \tiny$0.166511\times10^{51}$ & 
\tiny$0.166511\times10^{51}$ & \tiny$0.391088\times10^{12}$ 
& \tiny$0.123588\times10^{10}$  \\
\hline
\hline
\scriptsize $K^0$, $\overline{K}^{0}$---$d\overline{s}$, $\overline{d}s$ 
($K^0_L$--mode): & \scriptsize 
$r_0=<r>= 0.560964\ {\rm fm}$,  & \scriptsize $v_d= 0.999148 $, & 
\scriptsize $v_s= 0.984044 $ &  &\\
\hline 
\tiny $r$ & \tiny $n_{\rm coul}$ & \tiny $n_{\rm lin}$ 
& \tiny $n$ & \tiny $E$ & \tiny $H$ \\
\tiny (fm) & \tiny $ ({\rm m}^{-3}) $ 
& \tiny (${\rm m}^{-3}) $ & \tiny (${\rm m}^{-3}) $ 
& \tiny $({\rm V/m})$ & \tiny $({\rm T})$\\
\hline
\tiny $0.1r_0$ & \tiny $0.218299\times10^{67}$   
& \tiny $0.857308\times10^{65}$ & \tiny$0.226872\times10^{67}$ 
& \tiny $0.348022\times10^{24}$  & \tiny $0.116585\times10^{16}$ \\
\hline
\tiny$r_0$ & \tiny$0.218299\times10^{63}$ & \tiny$0.857308\times10^{63}$ 
& \tiny$0.107561\times10^{64}$& \tiny$0.348022\times10^{22}$  
& \tiny$0.116585\times10^{15}$  \\
\hline
\tiny$1.0$ & \tiny$0.216168\times10^{62}$  & \tiny$0.269778\times10^{63}$ 
& \tiny$0.291395\times10^{63}$ & \tiny$0.109516\times10^{22}$  
& \tiny$0.654000\times10^{14}$  \\
\hline
\tiny$10r_0$ & \tiny$0.218299\times10^{59}$  
& \tiny$0.857308\times10^{61}$ & \tiny$0.859491\times10^{61}$ 
& \tiny$0.348022\times10^{20}$  & \tiny$0.116585\times10^{14}$  \\
\hline
\tiny$a_0$ & \tiny$0.275668\times10^{43}$  & \tiny$0.963395\times10^{53}$ & 
\tiny$0.963395\times10^{53}$ & \tiny$0.391088\times10^{12}$ 
& \tiny$0.123588\times10^{10}$  \\
\hline
\end{tabular}
\end{center}
\end{table}

\section{Discussion and concluding remarks}
\subsection{Discussion}
 As is seen from Tables 3, 4, at the characteristic scales
of each meson the gluon concentrations are large and the corresponding fields 
(electric and magnetic colour ones) can be considered to be 
the classical ones with enormous strenghts. The part $n_{\rm coul}$ of gluon 
concentration $n$ connected with the Coulomb electric colour field is 
decreasing faster than $n_{\rm lin}$, the part of $n$ related to the linear 
magnetic colour field, and at large distances $n_{\rm lin}$ becomes dominant 
while quarks in mesons under investigation should be considered the 
ultrarelativistic point-like particles. It should be emphasized that in fact 
the gluon concentrations are much 
greater than the estimates given in Tables 3, 4 
because the latter are the estimates for maximal possible gluon frequencies, 
i.e. for maximal possible gluon impulses (under the concrete situation of 
pions and kaons). The latter also explains why gluon concentrations are 
much larger (about 8 orders of magnitude) for charged pions compared 
to $\pi^0$-meson: just the corresponding life times are different by the same 
orders so, accordingly, the conforming maximal gluon impulses are in 
inverse relation. 

The given picture is in concordance with the one obtained 
when considering charmonium in Refs. \cite{{Gon03},{GB04},{Gon052}}. 
As a result, the confinement mechanism developed in 
Refs. \cite{{Gon01},Gon051,Gon052} is confirmed by the considerations of the 
present Letter. 

It should be noted, however, that our results are of a preliminary character 
which is readily apparent, for example, from that the current quark masses 
(as well as the gauge coupling constant $g$) used in computation are known only within the 
certain limits and we can expect similar limits for the magnitudes 
discussed in the Letter so it is neccesary further specification of the 
parameters for the confining SU(3)-gluonic field 
in pions and kaons which can be obtained, for instance, by calculating decay 
constants and weak formfactors for the given mesons with the help of wave 
functions discussed above. Also one can obtain the analogous estimates for 
vector mesons, for instance, by computing the widths of radiative decays for 
them and so on. We hope to continue analysing other problems of meson 
spectroscopy elsewhere.

\subsection{Connection with potential approach and string-like picture}
The results obtained above allow us to shed some light on the following two 
problems. 
As is known, during a long time up to now in meson spectroscopy one often uses 
the so-called potential approach (see, e. g., Refs. \cite{{Rob},{Bra}} and 
references therein). The essence of it is in that the interaction between 
quarks is modelled on a nonrelativistic confining potential in the form 
$a/R+kR+c_0$ with some real constants $a, k, c_0$ and the distance between  
quarks $R$. On the other hand, also 
for a long time there exists the so-called string-like picture of quark
confinement but only at qualitative level (see, e. g., book of Perkins in 
\cite{RF}). Up to now, however, it is unknown as such 
considerations might be warranted from the point of view of QCD. Let us in short 
outline as our results (based on and derived from QCD-Lagrangian directly) 
naturally lead to possible justification of the mentioned directions. Thereto 
we note that one can calculate energy ${\cal E}$ of gluon condensate conforming 
to solution (4) in a volume $V$ through relation 
${\cal E}=\int_VT_{00}r^2\sin{\vartheta}dr d\vartheta d\varphi\>$ with $T_{00}$ 
of (16)--(17) but one should take into account that classical $T_{00}$ has a 
singularity along $z$-axis ($\vartheta=0,\pi$) and we have to introduce some 
angle $\vartheta_0$ (whose physical meaning is to be clarified a little below) 
so $\vartheta_0\leq\vartheta\leq\pi-\vartheta_0$. 
Then let us choose $V$ as shown in Fig. 1, i. e. the one between two concentric 
spheres with radii $R_0 < R$ restricted to interior of cone 
$\vartheta=\vartheta_0$. 
\begin{figure}
\vspace{0cm}
\caption{Integration domain for obtaining the gluonic energy}
\end{figure}

Without going into details (see also 
Ref. \cite{Gon051}) we shall obtain 
$${\cal E}={\cal E}(R)={\cal E}_0+\frac{a}{R}+kR, \eqno(20)$$
where  ${\cal E}_0=\frac{{4\pi\cal A}\cos{\vartheta_0}}{R_0}-
{2\pi\cal B}R_0\ln\frac{1+\cos{\vartheta_0}}{1-\cos{\vartheta_0}}$, 
$a=-{4\pi\cal A}\cos{\vartheta_0}$, $k={2\pi\cal B}\ln\frac{1+
\cos{\vartheta_0}}{1-\cos{\vartheta_0}}$ with constants ${\cal A},{\cal B}$ 
of (16). We recognize the mentioned confining potential in (20) when 
identifying ${\cal E}_0=c_0$ and we can see that phenomenological parameters 
$a, k, c_0$ of potential are expressed through more fundamental parameters 
connected with the unique exact solution (4) of Yang-Mills equations 
describing confinement. One can notice that the quantity 
$k$ (string tension) is usually related to the so-called Regge slope 
$\alpha^\prime=1/(2\pi k)$ and in many if not all of the papers using 
potential approach it is accepted $k\approx 0.18$ GeV$^2$, 
${\cal E}_0=c_0\approx-0.873$ GeV (see, e. g., Refs. \cite{{Rob},{Bra}}).  
Also one often uses parametrization $a=-4\alpha_s(R_0)/3$, where 
$\alpha_s(R_0)$ 
is the strong coupling constant at $R_0$ so when $R<R_0$ potential description 
is not applicable. If using (20) and the results obtained in Table 1 we can in 
series compute $\vartheta_0, \alpha_s(R_0), R_0$ for all mesons under discussion 
and also for 
the ground state of charmonium $\eta_c(1S)$ for that we use the parametrization 
from Refs. \cite{{GB04},{Gon051},{Gon052}} with replacing 
$a_i\to a_i/\sqrt{4\pi}, b_i\to b_i/\sqrt{4\pi}$ since the system of units in 
those references is different from the one used here. Results of computation 
are presented in Table 5. 

\begin{table}[htbp]
\caption{Parameters determining the confining potential for pions, kaons and 
charmonium ground state.}
\label{t.5}
\begin{center}
\begin{tabular}{|c|c|c|c|}
\hline
\small Particle & \small $\vartheta_0 $ & 
\small $\alpha_s(R_0)$  & $R_0$ (\small fm)\\
\hline
$\pi^{0}$---$\overline{u}u$  & \scriptsize $64.56^\circ$ & 
\scriptsize $0.676\times10^{-3}$ & 
\scriptsize 0.757  \\
\hline
$\pi^{0}$---$\overline{d}d$  & \scriptsize $64.276^\circ$ & 
\scriptsize $0.532\times10^{-3}$ & 
\scriptsize 0.957 \\
\hline
$\pi^{\pm}$---$u\overline{d}$, $\overline{u}d$ 
  & \scriptsize $64.199^\circ$ & \scriptsize $0.600\times10^{-3}$ & 
\scriptsize 0.957 \\
\hline
$K^{\pm}$---$u\overline{s}$, $\overline{u}s$ 
  & \scriptsize $82.464^\circ$ & \scriptsize $0.357\times10^{-2}$ & 
\scriptsize 0.958 \\
\hline
$K^{0}$, $\overline{K}^{0}$---$d\overline{s}$, $\overline{d}s$ 
  & \scriptsize $60.272^\circ$ & \scriptsize $0.986\times10^{-2}$ & 
\scriptsize 0.960 \\
\hline
$\eta_c(1S)$---$c\overline{c}$ 
  & \scriptsize $89.934^\circ$ & \scriptsize $ 0.163\times10^{-2}$ & 
\scriptsize 0.957 \\
\hline
\end{tabular}
\end{center}
\end{table}
As well as in Ref. \cite{Gon051}, we may consider $\vartheta_0$ to be 
a parameter determining some cone $\vartheta=\vartheta_0$ so the quark  
emits gluons outside of the cone and, generally speaking, the angle 
$\vartheta_0$ is increasing with quark mass. We can see that though a 
potential could be associated 
with each meson but the potential is invalid at the meson characterisic 
scales which follows from Table 2. Moreover, as was shown in 
Refs. \cite{{Gon051},{Gon052}}, potential of form (20) cannot be a solution 
of the Yang-Mills equations if simultaneously $a\ne0, k\ne0$. Therefore, 
it is impossible to obtain compatible solutions of the 
Yang-Mills-Dirac (Pauli, Schr{\"o}dinger) system when inserting potential of form 
(20) into Dirac (Pauli, Schr{\"o}dinger) equation. So, we draw the conclusion 
(mentioned as far back as in Refs. \cite{Gon03}) that the potential approach 
seems to be inconsistent. 

Now if there are two quarks $Q_1, Q_2$ and each of them emits gluons outside 
of its own cone $\vartheta=\vartheta_{1,2}$ (see Fig. 2) then we have 
soft gluons 
(as mentioned in Section 1) in regions I, II and between quarks so 
a characteristic transverse size $D$ of the gluon condensate is decreasing 
with increasing quark masses as we just now saw. For heavy quarks the gluon 
configuration between them practically transforms into a string. As a result, 
there arises the string-like picture of quark confinement but the latter seems 
to be warranted enough only for heavy quarks. It should be emphasized that 
string tension $k$ is determined just by parameters $b_{1,2}$ of linear 
magnetic colour field from solution (4) [see (20)] which indirectly confirms 
the dominant role of the mentioned field for confinement. 

\subsection{Concluding remarks}
Considerations of the present Letter as well as ones of 
Refs. \cite{{Gon03},{GB04}} show that in meson spectroscopy the approach 
based on the unique family of compatible nonperturbative solutions for the 
Dirac-Yang-Mills system 
derived from QCD-Lagrangian may be employed for both light mesons and heavy 
quarkonia. Under the circumstances there are no apparent obstacles to 
apply the approach to any meson. We hope further studies along this direction 
to confirm the given point of view. 

\ack

    Author is grateful to the referee whose valuable remarks promoted 
essential improvement of the initial version of the Letter.

\begin{figure}
\vspace{0cm}
\caption{Formation of string-like picture between quarks}
\end{figure}


\end{document}